\documentclass[fleqn,10pt]{wlscirep}
\usepackage[utf8]{inputenc}
\usepackage[T1]{fontenc}
\usepackage{multirow}
\usepackage{lineno}
\usepackage{setspace}

\usepackage[colorlinks=true, allcolors=blue]{hyperref}
\title{Revealing configurational attractors in the evolution of modern Australian and US cities}

\author[1,*]{Bohdan Slavko}
\author[1]{Kirill S. Glavatskiy}
\author[1]{Mikhail Prokopenko}
\affil[1]{Centre for Complex Systems, The University of Sydney, Sydney, NSW 2006, Australia}

\affil[*]{To whom correspondence should be addressed: bohdan.slavko@sydney.edu.au.}


\begin{abstract}
The spatial structure of modern cities exhibits highly diverse patterns and keeps evolving under numerous constraints and sustainability demands. Two key dimensions have recently achieved prominence in characterizing this diversity: heterogeneity and spreading. However, modern settlements do not fill the entire heterogeneity--spreading space. Yet, the dynamic mechanisms leading to emergence of the observed layouts are unclear. Here, we assess the heterogeneity and spreading of population density in 25 Australian and 175 US cities. We observe that larger cities tend to form a cluster with a low degree of spreading and a high degree of heterogeneity, and relate this observation to the dynamic properties of intra-urban migration in these cities. In doing so, we introduce a model consistent with the relocation data which predicts such highly compact and heterogeneous structure for the majority of cities, in concordance with the actual layout data. In addition, we analyze the stability of the long-term dynamics of urban configurations with respect to changes in the mobility characteristics, such as social disposition and relocation impedance near their equilibrium states. As a result, we report three qualitatively distinct feasible phases of urban structures: uniform, monocentric, and polycentric. These phases are shown to be separated by either smooth or sharp transitions, observed in the space of suitably chosen configurational parameters. Finally, this analysis reveals that the set of all possible equilibrium configurations (``configurational attractors'') form a narrow region in the heterogeneity--spreading space, thus explaining the emergence of sustainable clustering patterns.
\end{abstract}
\begin{document}

\flushbottom
\maketitle
%
%
\thispagestyle{empty}


\onehalfspacing

\noindent Spatial distribution of population in large cities affects various aspects of human well-being, such as economic productivity (more compact cities reduce overhead business costs) and ecological footprint (home-to-work travel by public and private transport increases greenhouse gas emissions).
Spatial structure of cities varies with respect to population distribution \cite{Fujita1982, Anderson1996, Mcmillen2001, Bertaud2003, Green2007, Louf2013, Louail2014}, ranging from monocentric, with major human activities localised around the central district, to polycentric, with multiple centres of residential land use and their business activity. Each particular structure affects the dynamics of human activities \cite{Meijers2010, Fang2015} and can also be seen as a stationary configuration produced by associated economic mechanisms (e.g., increasing return to scale, transportation cost minimisation, market of rents and wages) \cite{Fujita1982, Louf2013, Wilson2008,Osawa2017, Crosato2018}. The currently adopted models  relate the urban layout to human activities but do not typically provide a 
dynamic mechanism explaining how sustainable spatial configurations can be reached  \cite{Louf2013, Crosato2018, Strano2020, Crosato2020}.  Development of a consistent quantitative framework describing mechanisms which shape cities into feasible configurations remains a formidable challenge~\cite{Fujita2010, Batty2013, Barthelemy2016book}. 


The lack of consensus in rigorously quantifying the urban structure is emphasized by the multiplicity of indirect indicators adopted in recent literature \cite{Galster2001,Bettencourt2010, Louf2013, Louf2014scaling, Arcaute2015,Sarkar2018, Sahasranaman2019}. 
Most recently, two contenders for a concise quantification of different urban structures have emerged in order to separate two orthogonal dimensions: heterogeneity (varying from highly homogeneous to nonuniform asymmetric patterns) and spreading (varying from compact to sprawled configurations)~\cite{Tsai2005,Louail2014,Volpati2018}. To a significant extent, these dimensions subsume several prior characteristics. 
For example, the sprawl or degree of compactness (i.e, spreading) has been previously measured as the average distance to the central business district \cite{Bertaud1999}, the population density gradient \cite{Bertaud2003}, a degree of density clustering (e.g., Moran's index and Geary's coefficient) \cite{Galster2001,Tsai2005}, or the average distance between population hot-spots \cite{Louail2014,Volpati2018}. The introduction of the heterogeneity index~\cite{Volpati2018} was needed because the  spreading characteristics alone may not be sufficient to fully characterize the spatial structure of a city, as the same value of the degree of spreading may be obtained for two qualitative different urban configurations, e.g. a city with the uniformly distributed population density and a polycentric city \cite{Tsai2005}. The heterogeneity index is related to the Gini coefficient \cite{Tsai2005, Volpati2018}, the entropy, or the relative standard deviation \cite{Volpati2018}. 

\begin{figure*}
\centering
\includegraphics[width=17.8cm]{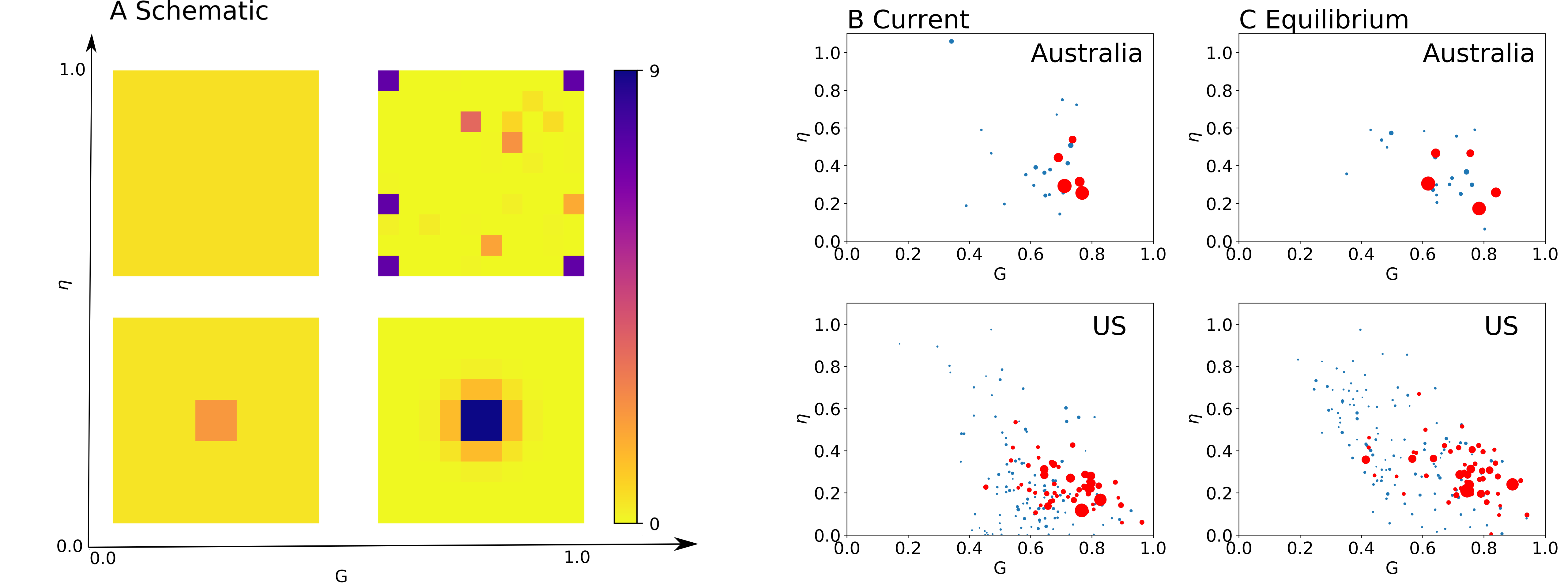}
\caption{The layout diagram for spreading index $\eta$ and heterogeneity index $G$. (A) Schematic representations of the city layout for extreme values of the heterogeneity and spreading indicators: homogeneous and disperse ($G=0.0018$, $\eta=0.9704$; upper left), heterogeneous and disperse ($G=0.9139$, $\eta=1.2435$; upper right), homogeneous and compact ($G=0.1646$, $\eta=0.1324$; lower left),  heterogeneous and compact ($G=0.9079$, $\eta=0.1646$; lower right). The centre(s) are compact and highly populated, as indicated by the blue color, while the population density in the remaining area is represented by various tones of yellow. Arrangement of the Australian and US urban areas in the heterogeneity--spreading space: current (B) and near equilibrium (C). The size of each circle corresponds to the size of the corresponding urban area; urban areas with population above 1 million people are shown in red; for presentation purpose, Australian and US cities are shown in different scale. }
\label{fig:spreading_heterogeneity_diagram}
\end{figure*}

Thus, the joint heterogeneity--spreading space can concisely represent a variety of spatial layouts: monocentric, polycentric, and sprawled, as well as their combinations.  Fig.~\ref{fig:spreading_heterogeneity_diagram}A shows schematic layouts when these indicators vary between 0 and 1. In the two-dimensional heterogeneity--spreading space (referred to as the layout diagram), any urban area (i.e. a city) is characterized by a point, and the closeness of two points reflects the similarity of the corresponding spatial urban layouts. 

It is remarkable that the layout diagrams produced for sufficiently large samples (e.g., comprising all cities from  fairly populous countries) are not homogeneously populated. The unoccupied regions of the space indicate that the corresponding layouts are infeasible as sustainable urban configurations. 
This suggests that the emergence of specific layouts is driven by some underlying dynamic mechanisms shaping the urban structure over time, as a result of intricate intra-urban migration flows. 
It was recently shown that cities in France tend to cluster along a specific region of the layout diagram (the cross-diagonal connecting the low heterogeneity/high spreading region with the high heterogeneity/low spreading region~\cite{Volpati2018}). We performed a similar analysis for the cities in Australia and USA, and also found a distinct pattern, emphasizing the tendency of larger cities to exhibit high heterogeneity/low spreading (i.e., monocentricity), as shown in Fig.~\ref{fig:spreading_heterogeneity_diagram}B. A particular location of the  layout cluster observed for the USA and Australia is different from the one reported for France. This difference may be attributed to the divergence  between historical trajectories of the European settlements  on the one hand, and those of the American and Australian cities on the other hand: the latter trajectories have been less restricted by space constraints and thus, resulted in lower population density than the former. However, the  existence of clustering in the heterogeneity--spreading space supports our conjecture that specific relocation dynamics drive the formation of urban structures in terms of the population density.

We verify this conjecture by explicitly relating the population distribution, which determines the spatial urban layout, to the long-term migration flows, driven by multiple factors. 
We focus on two generic factors: the relative location attractiveness and the geographic proximity. The former factor represents tendency to migrate from less to more attractive locations \cite{Harris1978, Wilson2008, Crosato2018}, with the degree of attractiveness dependent on qualities of local infrastructure, as well as other location characteristics \cite{Mcfadden1978, Kim2005, Perez2003, Wu2019, Crosato2018, Dynamic_resettlement_paper, Slavko2020}. The latter factor captures that the migration flow is the most intensive when its origin and destination are close to each other, reflecting human aversion to lifestyle changes \cite{Slavko2020}. Such distance-dependent relationships are also observed in a broad class of economic and social activities distributed in space: human mobility \cite{Zipf1946, Reilly1929, Huff1964, Harris1978, Wilson2008}, supply chains and economic interactions~\cite{Zipf1946, Jung2008, Kaluza2010}, initially motivated as the ``gravity law" \cite{Zipf1946, Huff1964}, and subsequently formalized in various ``gravity'' models. 

In order to quantify the dependence between residential migration flows and their effects on the spatial urban layouts, we propose a dynamic model that describes resettlement as a process driven by the location attractiveness and geographic proximity. This model is validated with  short-term internal migration data, producing consistent predictions for different input datasets (i.e., Australian Census data 2011 and 2016). Using this model we achieve several objectives. Firstly, we demonstrate that there exist three distinct phases into which cities evolve: (i) a uniform phase where all suburbs have similar population density (with low heterogeneity and large spreading), (ii) a monocentric phase comprising a single center of high population density (medium to large heterogeneity and low spreading), and (iii) a polycentric phase with multiple centers (large heterogeneity and medium spreading). This provides a link between the microscopic relocation dynamics and the macroscopic state of urban configurations. Secondly, we show that  only a subset of possible city layouts is feasible in practice: these layouts correspond to dynamical ``configurational'' attractors. This, in turn, explains the observed clustering patterns on the layout diagram.
Furthermore, we assess spatial configurations of 25 Australian and 175 American metropolitan areas, and show that our model is capable of reproducing the patterns observed in these cities: specifically, the configurations predicted by our model form a cluster of cities with high degrees of heterogeneity and compactness, in concordance with actual observations.

\section*{Microscopic urban mobility model}

We describe the migration flow between two suburbs as being driven by a spatial difference in the residential potential. This approach is inspired by the analogy with diffusion, in which the matter flow is driven by a spatial difference in the chemical potential \cite{Slavko2020b, deGrootMazur}, and adopts the master equation approach introduced by Weidlich \textit{et al} \cite{Haag1984,Weidlich1990} to describe settlement formation, extended by the spatial interaction term \cite{Slavko2020}.

Let $\tau_{ij}$ be the migration flow density, i.e. the number of people that relocate from the source suburb $i$ to the destination suburb $j$ per unit area of the destination suburb within the unit of time. Furthermore, let $x_i$ be the population and $u_i$ be the residential potential (attractiveness) of suburb $i$. If the distance between the source and the destination suburbs is $d_{ij}$, then the migration flow between them is described as  
\begin{equation}\label{eq:migration_flow}
    \tau_{ij} = \mu\, x_i\, e^{\alpha (u_j-u_i) -\gamma\,d_{ij}}
\end{equation}
where $\mu$, $\alpha$ and $\gamma$ are the parameters which characterize the entire urban area and are therefore identical across all suburbs. In particular, $\alpha$ is the \textit{social disposition}, which quantifies importance of the attractiveness gain when moving between suburbs; $\gamma$ is the \textit{relocation impedance}, which indicates a degree of aversion to long-distance relocation; $\mu$ is a normalizing scale factor. Given transience of residential relocations, the population $x_i(t)$, the migration flow $\tau_{ij}(t)$, and the residential potential $u_i(t)$ are time-dependent. In contrast, the parameters $\mu$, $\alpha$, $\gamma$ are the characteristics of the urban area as a whole and do not change with time. 

The residential potential may in principle depend on various characteristics, such as quality of local infrastructure (schools and other local services) \cite{Crosato2018, Dynamic_resettlement_paper}, distance to the city center \cite{Kim2005, Perez2003}, etc., and may also be inferred directly from data \cite{Slavko2020}. In this study we express the residential potential as a non-monotonic function \cite{Dynamic_resettlement_paper, Haag1984} of the population density, $\rho$:
\begin{equation}
    u(\rho)=\rho \left(1-\frac{1}{2}\frac{\rho}{\rho_0}\right).
\end{equation}
where $\rho_0$ is a city-specific parameter. This expression can be interpreted as if the attractiveness of a suburb increases with its density (people tend to avoid both underpopulated areas expecting low quality of public infrastructure due to insufficient demand) until it reaches a certain saturation threshold $\rho_0$, beyond which attractiveness starts decreasing (people tend to avoid overpopulated areas expecting low quality of public infrastructure due to insufficient capacity).


Using this model we analyse the structure of Australian and US cities in the long-run, by simulating their evolution trajectories for the next 100 years according to the algorithm provided in \textit{Methods}. 
The model parameters are estimated through calibration to the migration data. The calibration procedure for Australian cities is reported in \textit{Methods}. The accuracy of the parameter estimates is presented in Table S1 in SI. The US census does not keep track of previous places of residence and, therefore, we interpolate the parameters values as described in \textit{Methods}.

Figure~\ref{fig:spreading_heterogeneity_diagram} contrasts actual configurations of the considered cities (B) and their simulated long-run counterparts (C) in the heterogeneity--spreading space. Each representation exhibits distinct clustering of the layouts, and the comparison highlights a strong similarity between the clusters formed by actual and long-run configurations. Moreover, there is a clear tendency for larger cities to exhibit more heterogeneous and compact layouts. When social disposition and relocation impedance are maintained, the future long-term migration dynamics towards the equilibrium configurations are unlikely to produce configurations much different from the currently observed. In other words, the current state of affairs appears to be close to the equilibrium, with larger cities already reaching their stable layouts.


\section*{Three phases of cities}

\begin{figure*}
\includegraphics[width=17.8cm]{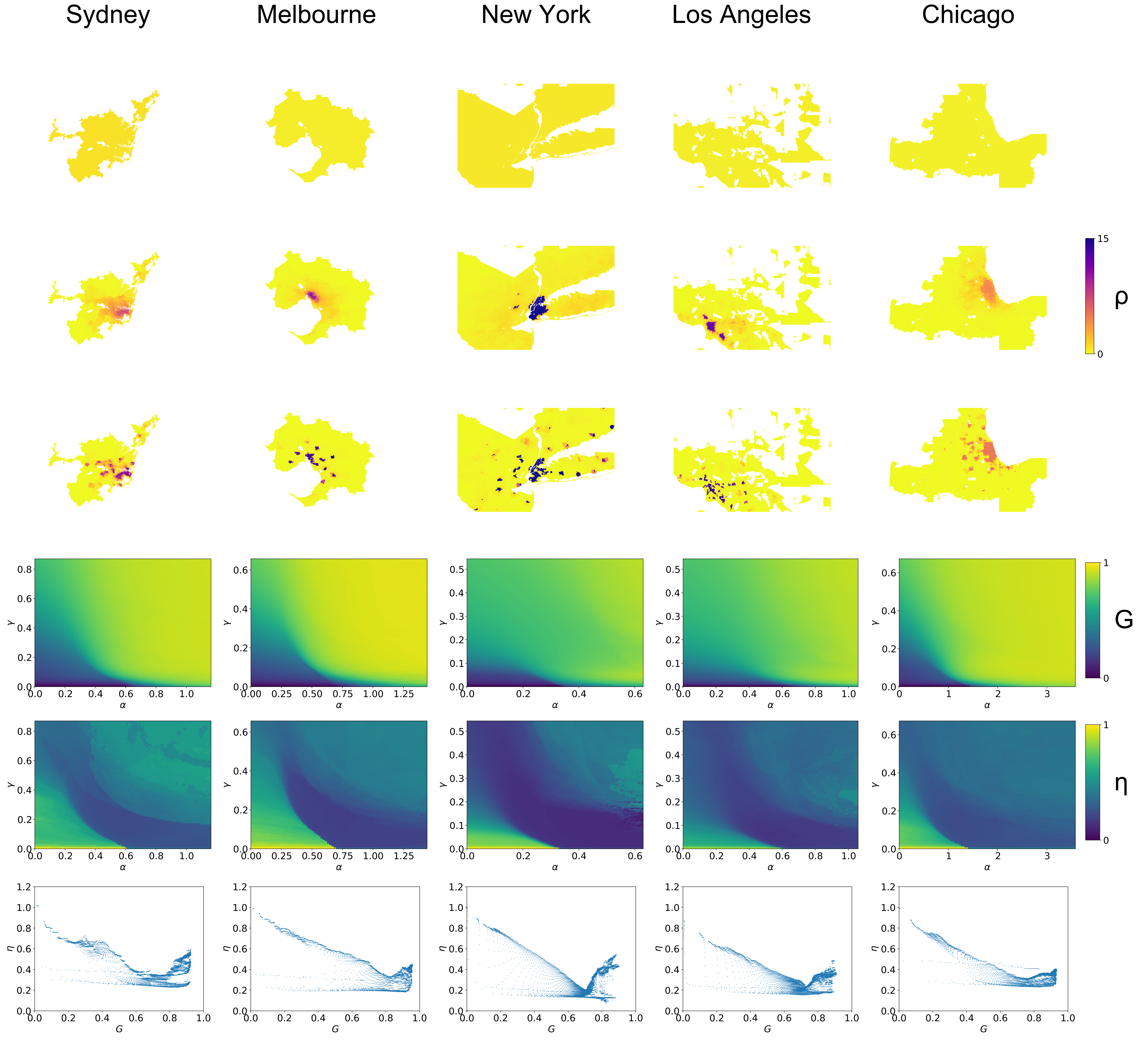}
\caption{Equilibrium phases of major Australian and US cities. We show the population density maps for three distinct phases of equilibrium city structure: uniform (first row), monocentric (second row) and polycentric (third row).  The values of $G$ (fourth row) and $\eta$ (fifth row) near equilibrium are given as functions of $\alpha$ and $\gamma$, with their values also plotted against each other (in $(G,\eta)$ space; sixth row). Values of the population density are measured in thousands of residents per $km^2$. Values of $\alpha$ and $\gamma$ are divided by the factor of $10^{-3}$.}
\label{fig:joint_phase_diagrams}

\end{figure*}

To explore other possible configurations of the cities
we trace our prediction for different values of the model parameters ($\alpha$ and $\gamma$).  
The colormaps in Fig.~\ref{fig:joint_phase_diagrams} show the values of heterogeneity and spreading indices for the largest Australian and US cities. From this diagrams, we observe that all major cities have three qualitatively distinct phases. The first one has a low degree of heterogeneity and a high degree of spreading, characterizing the uniform phase (the first row in Fig.~\ref{fig:joint_phase_diagrams}) with low values of $\alpha$ and $\gamma$. The second phase has a high degree of heterogeneity and the lowest degree of spreading, describing the monocentric phase with the medium values of $\alpha$ and $\gamma$ (the second row in Fig.~\ref{fig:joint_phase_diagrams}). The last one has a high degree of heterogeneity and a medium degree of spreading, characterizing the polycentric phase (the third row in Fig.~\ref{fig:joint_phase_diagrams}) with large values of $\alpha$ and $\gamma$. 

These results are quite intuitive. If $\alpha=0$, $\gamma=0$, there is no spatial heterogeneity and all suburbs are equally attractive for their residents. Hence, the city moves towards a structure with homogeneous population density  (first row in Fig.~\ref{fig:joint_phase_diagrams}), characterized by zero value of the heterogeneity index and unit value of the spreading index. As $\alpha$ and $\gamma$ increase, the central districts become more populated and the city becomes clearly monocentric (second row in Fig.~\ref{fig:joint_phase_diagrams}). This takes place due to two reasons. Firstly, the central suburbs are more populated in the beginning which makes them more attractive in terms of $u$. Hence, an increase in $\alpha$ motivates people to relocate towards the central areas. Secondly, the central central districts are closer to all other suburbs (on average) than the peripheral ones. 
It can be shown from the detailed balance condition that even in the case where $\alpha=0$, the population density in suburb $i$ is proportional to the average proximity to the other suburbs:
$\rho_i \propto \sum_{j \neq i} e^{-\gamma d_{ij}}S_j$.
This means that a better average proximity to other suburbs generates a higher migration inflow to the central districts as $\gamma$ increases.  

If the social disposition and relocation impedance are very large, central areas pass the saturation threshold and their attractiveness starts decreasing. Meanwhile, non-central clusters of high population density (the suburbs that have a lower population density than the central districts but a higher density than the peripheral ones) appear more competitive than the central districts. Consequently, these non-central clusters attract a significant part of residents from the surrounding areas. This results in the emergence of multiple local clusters which are well-separated from the central one (third row in Fig.~\ref{fig:joint_phase_diagrams}) and shape the city towards a polycentric structure. 

Transitions between these phases can be  smooth or abrupt. The examples for Sydney and Melbourne are shown in Fig.~S1-S2 in SI. In particular, if we vary $\gamma$ (for the fixed level of $\alpha$), Sydney passes from the uniform to the monocentric configuration rather smoothly. However, an analogous transition in Melbourne is abrupt: a pronounced center forms only when $\gamma$ reaches the level of 0.646, with the configuration being relatively uniform below this threshold. The abruptness of this transition is caused by existence of multiple stable equilibria which form when $\alpha$ is large. Depending on the initial population structure and the model parameters, the system may converge to one of these equilibria, and then switch between them, producing a discontinuous transition. This reinforces dynamic resettlement as a mechanism of phase transitions in urban configurations~\cite{Haag1984,Dynamic_resettlement_paper}. 


\section*{Configurational attractors: evolution of urban layouts}

To re-iterate, each urban area can be characterized by spreading index $\eta$ and heterogeneity index $G$. The resultant phase diagrams constructed in terms of the control parameters $\alpha$ and $\gamma$ allow us to explore the long-term dynamics in the space of these indicators. As these indicators cover a wide range of actual spatial layouts (see Fig.~\ref{fig:spreading_heterogeneity_diagram}A), a set of attractors in this space describes a specific range  of possible spatial layouts for each particular city. While $\alpha$ and $\gamma$ are the parameters of the microscopic model capturing individual preferences of residents, 
the urban layout indicators $\eta$ and $G$ measure the actual spatial configuration of an entire urban area. This enables a macroscopic analysis of emergent urban layouts. 

Feasible equilibrium configurations across several cities are shown in  Fig.~\ref{fig:joint_phase_diagrams} (last row). For each city, it is evident that  not all potential configurations can be attained. In fact, there exists only a narrow configurational region in the $(G,\eta)$ space. This regions represents a set of configurational attractors, the equilibrium points of the evolution of the dynamical system described by \eqref{eq:migration_flow}. The shape of the region in the $(G,\eta)$ space represents the layout ``signature'' of the city.

It is remarkable that there is a well-defined recurrent motif across the  signatures of major urban areas. 
In particular, the spatial layout of each city varies between being heterogeneous and compact, on the one hand, and being homogeneous and dispersed, on the other hand, while following a distinct profile $\eta(G)$ connecting the upper-left and lower-right corners. In the upper-left corner, the degree of scattering is low, but it gradually increases as $G$ approaches 1. In contrast, the configurations with high heterogeneity and spreading (upper-right) are not feasible. The layouts with a homogeneous and compact structure (lower-left) are  possible: in terms of our microscopic model, this corresponds to a low relocation impedance, i.e., the situation when people move easily to a more attractive area irrespectively of its location. This may, in particular, be the case if the city is new and most of the residents have moved in recently, and have not yet established strong connections to some preferred geographical locations. 

The emergence of feasible urban layouts is in a good agreement with the  data on  existing urban configurations. In particular, we do not observe compact and homogeneous, or dispersed and heterogeneous cities, see  Fig.~\ref{fig:spreading_heterogeneity_diagram}B. Previous studies also did not report such layouts~\cite{Volpati2018}.
The feasible region with the highest concentration of attractors is shaped by the higher values of heterogeneity $G$ and lower values of spreading $\eta$ (see the last row of Fig.~\ref{fig:joint_phase_diagrams}). This suggests that the most representative urban layout is a monocentric city. This, in turn, explains compact clustering of the actual US and Australian cities shown in Fig.~\ref{fig:spreading_heterogeneity_diagram}. In other words, the recurrent signature observed in the heterogeneity-spreading space suggests a tendency for actual cities to evolve towards these monocentric layouts, dominating the diversity of initial conditions. 

This tendency is not universal, being observed only in the largest cities. This conclusion is in concordance with the study of French cities~\cite{Volpati2018} which included a significant number of small cities, and as consequence, reported a much wider cluster than the cluster formed by the US or Australian cities described in our study. Nevertheless, both clusters have a well-defined profile in the heterogeneity-spreading space.  
The feasibility of specific spatial structures depends on the current urban configuration, as well as a historical trajectory of the city's evolution. For example, highly heterogeneous or highly compact configurations do not appear feasible for a relatively small city, such as Canberra, even in the long-run (as shown in Fig.~S3 in SI). The urban layout for a particular city and its current position on the urban layout diagram may also be influenced by the city evolution, constrained by various limits and planning decisions. For instance, Canberra was mostly planned in a centralized fashion, and its short modern history did not include significant self-organized resettlement flows. Consequently, the relative extents of  the  heterogeneity and spreading exhibited by resultant attractors should reveal and contextualize unique features of the urban evolution.


\section*{Conclusions}

We have explained the macroscopic patterns in  spatial layouts of American and Australian cities by their microscopic dynamics of residential migration. In doing so, we introduced a model that is consistent with the short-term intra-urban migration data and is capable of explaining high degrees of heterogeneity and compactness observed in the majority of the considered American and Australian cities.
Importantly, our model reveals three qualitatively distinct equilibrium phases: uniform, monocentric and polycentric. The first one is driven only by the mobility parameters. 
In the other two phases, the long-term city structure depends on both the initial configuration and the model parameters, and the transitions between the phases may be sharp. 

Our predictions enhance the previous research which quantified possible configurational transitions in Australian cities \cite{Crosato2018, Slavko2020, Crosato2020, Slavko2020b}. This is significant, given that the prior research used different types of data, distinct from the residential migration data, and different model types (i.e., not based on a microscopic master equation). The results presented in our work reinforce the conclusion that critical regimes in urban dynamics 
are inherent to a wide class of models \cite{Crosato2018,Dynamic_resettlement_paper,Penny2018demise}

An important difference emphasized in our study is that the monocentric or polycentric structures emerge in response to changes in social disposition and relocation impedance, rather than optimisation of the commuting cost, departing from several canonical models \cite{Fujita1982,Fujita2010,Wilson2008,Louf2013,Crosato2018}. This indicates that cities can be monocentric or polycentric even if when the commuting costs  have a low importance for the city residents.


The presented approach contributes to a quantitative framework that brings together spatial and dynamic properties of cities, inspired by physics-based notions of thermodynamic fluxes and phase transitions. It could open a way to tracing the historical  trajectories within a constrained space. In particular, the emergence of specific city layouts,  quantified along the  heterogeneity and spreading axes, can be related to specific attractors attainable within a dynamic model. 

Our results can be useful for urban planners and other practitioners. Some urban configurations are not attainable: the cities are unlikely to shape as homogeneous and compact, or heterogeneous and dispersed, in response to sustainability constraints. In addition, the initial conditions may limit the variety of the configurations achievable in the long-run due to the path-dependency of sustainable urban evolution. Thus, short-term development projects need to be consistent with the long-term goals which have long-lasting impact on the feasibility of  preferred city configurations.

\section*{Methods}

\subsection*{Spatial indices}
We adopt the methodology introduced in \cite{Louail2014,Volpati2018} and quantify the spatial structure of cities by degrees of heterogeneity and spreading. 
The heterogeneity index $G$ is defined as 
\begin{equation}
    G = \frac{\sum_{i=1,j=1}^{N} |\rho_i - \rho_j| S_i S_j}{2 \sum_{i=1}^{N} \rho_i S_i \sum_{j=1}^{N} S_j},
\end{equation}
where $\rho_i$ is the population density in suburb $i$, and $S_i$ is the corresponding area.
The spreading index $\eta$ is given by:
\begin{equation}\label{spreading_index}
    \eta\, =\, \frac{\sum_{i,j}d_{ij}\,\xi_i(\overline{\rho})\,\xi_j(\overline{\rho})}{\sum_{i,j}d_{ij}S_i S_j}
\end{equation}
where $\xi_i(\overline{\rho}) \equiv S_i \Theta(\rho_i - \overline{\rho})\,S/S(\rho)$, $\overline{\rho}$ is the LouBar threshold, defined according to the method described in \cite{Louail2014,Volpati2018}, 
$\Theta(x)$ is the Heaviside step function, $S_i$ is the area of suburb $i$, $S$ is the total area comprising all suburbs, and $S(\rho)$ is the area of suburbs with the population density higher than $\rho$.

\subsection*{Model calibration}
Our analysis is performed for 25 Australian and 175 US cities. Model parameters are calibrated to the five-year internal migration data set published by the Australian Bureau of Statistics, using the Statistical Area Level 2 (SA2) \cite{ABS2016} resolution. 
A fitting procedure is conducted numerically by solving the residual minimisation problem: $\hat{\theta}=\arg \min_{\theta}\sum_{i,j=1}^{N}(T_{ij}(\theta)-T_{ij})^2$,
where $\theta=(\alpha,\gamma,\rho_0,\mu)$ is a vector of parameters, $T_{ij}$ is the actual migration flow from area $i$ to area $j$, and $T_{ij}(\theta)$ is the quantity calculated through \eqref{eq:migration_flow}. The information about previous places of residence  is not available in the US census dataset. For US cities we use only population density data in ZIP Code Tabulation Area (ZCTA) resolution \cite{USCB2016}.
To obtain the parameters for the US cities, we interpolate them linearly over the grid of population density. We use the LouBar population density as a representative value for each city. Some of the US LouBar density values are higher than those in all Australian cities. For these cities we use the median value for parameter $\gamma$ (there is no obvious trend in values of $\gamma$ as shown in Fig.~S4 in SI) and use log-linear extrapolation to obtain values of $\alpha$ and $\rho_0$ (as shown in Fig.~S4 in SI). 


\subsection*{Simulation}
Simulation of the population vector $x(t)=(x_1(t),x_2(t),\dots x_N(t))$ describing population in each area is done through the following difference equation:
\begin{equation}\label{eq:dynamic_equation}
    x(t+1)=x(t)P(t)
\end{equation}
where $P(t)$ is a migration rate matrix where each element $p_{ij}$ is a fraction of people moving from $i$ to $j$ among total population of suburb $i$.
These migration rates are given by:
\begin{equation*}
    p_{ij}=
    \begin{cases}
    \varepsilon h_{ij}, \text{ if } i \neq j\\
    1-\varepsilon, \text{ if } i = j,
    \end{cases}
\end{equation*}
where $\varepsilon$ is a mobility parameter defining proportion of people who change their place of residence within one-year period. 
It has been shown previously \cite{Slavko2020b} that a resettlement model is inconsistent with the dynamic data if the relocation rate $\varepsilon$ is homogeneous across the whole population. Consequently, at least two groups with different $\varepsilon$ are necessary to reproduce the dynamic pattern present in the data. Hence, we assume existence of two heterogeneous groups, and calibrate the corresponding values of $\varepsilon$ according to the technique described in \cite{Slavko2020b}.
Values $h_{ij}$ are defined as follows:
\begin{equation}
    h_{ij}(t)=\frac{\tau_{ij}(t)S_j}{\sum_{k\neq i}\tau_{ik}(t)S_k}.
\end{equation}
To predict the city structure near the long-run equilibrium, we iterate \eqref{eq:dynamic_equation} 100 times, as matrix $P$ depends on the population structure $x(t)$ and changes over time. A 100-year forecast is chosen as a reasonable approximation of the equilibrium state. All computations are done in Python 3.7.3.

\section*{Acknowledgments}
\textbf{Data accessibility.} All data needed to evaluate the conclusions in the paper are available from Australian Bureau of Statistics \cite{ABS2016}.\\
\textbf{Authors’ contributions.} BS analyzed the data, all authors developed the model and prepared the manuscript.\\
\textbf{Competing interests.} The authors declare that they have no competing interests.\\
\textbf{Funding.} This work is supported by the University of Sydney's Postgraduate Research Scholarship SC0789, and the Australian Research Council Discovery Project DP170102927.

\bibliography{urban-attractor-bibliography}

\newpage

\section*{Supplementary information}
\setcounter{figure}{0}
\setcounter{table}{0}

\begin{figure}[ht]
\centering
\includegraphics[width=.9\linewidth]{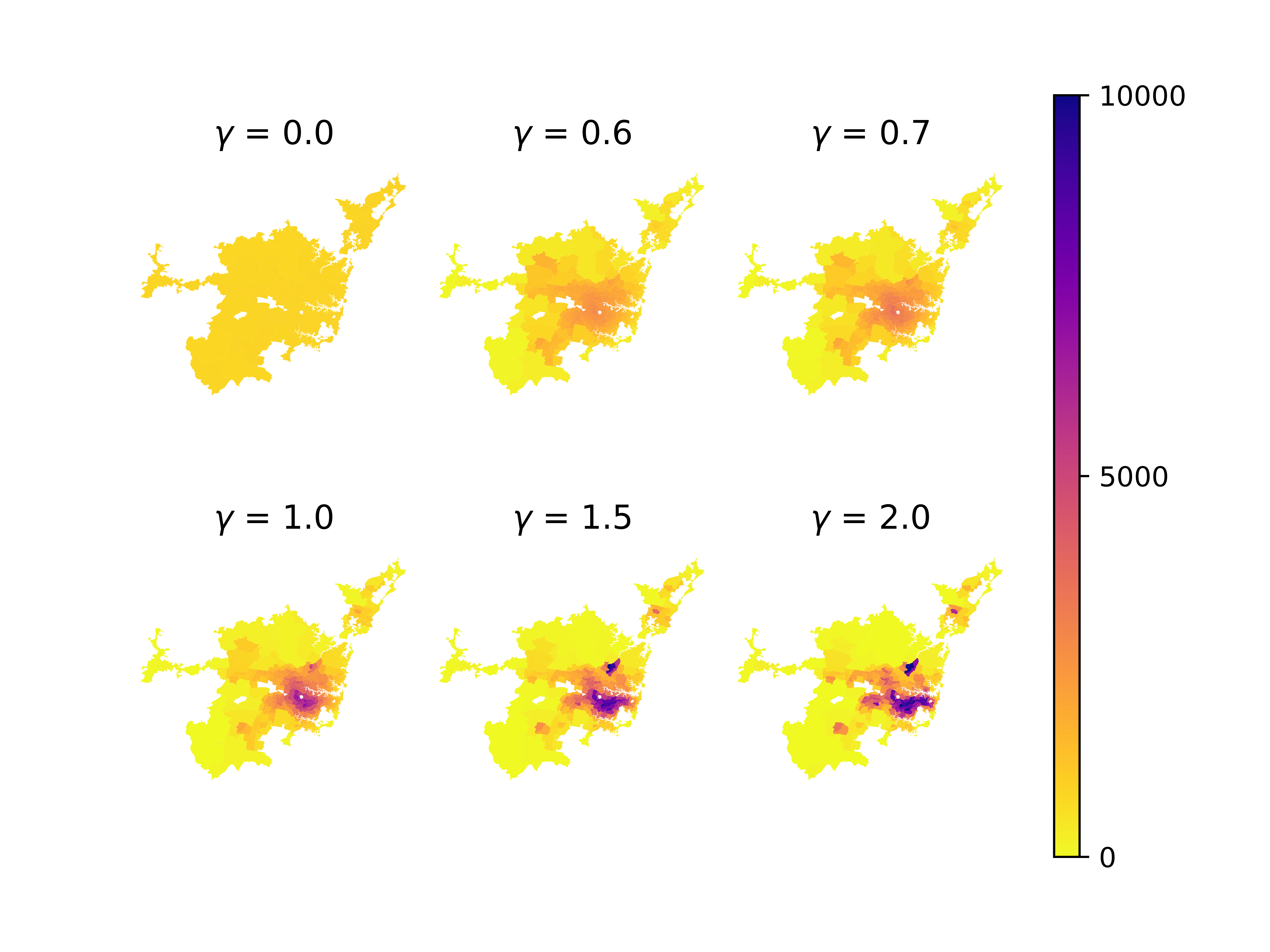}
\caption{Smooth transition between uniform, monocentric and polycentric phases in Sydney: population density maps for different values of $\gamma$. Value of $\alpha$ is fixed at the current level. Values of $\gamma$ are shown in relative units: divided by the estimated current value.}
\label{fig:Sydney_smooth}
\end{figure}

\begin{figure}
\centering
\includegraphics[width=.9\linewidth]{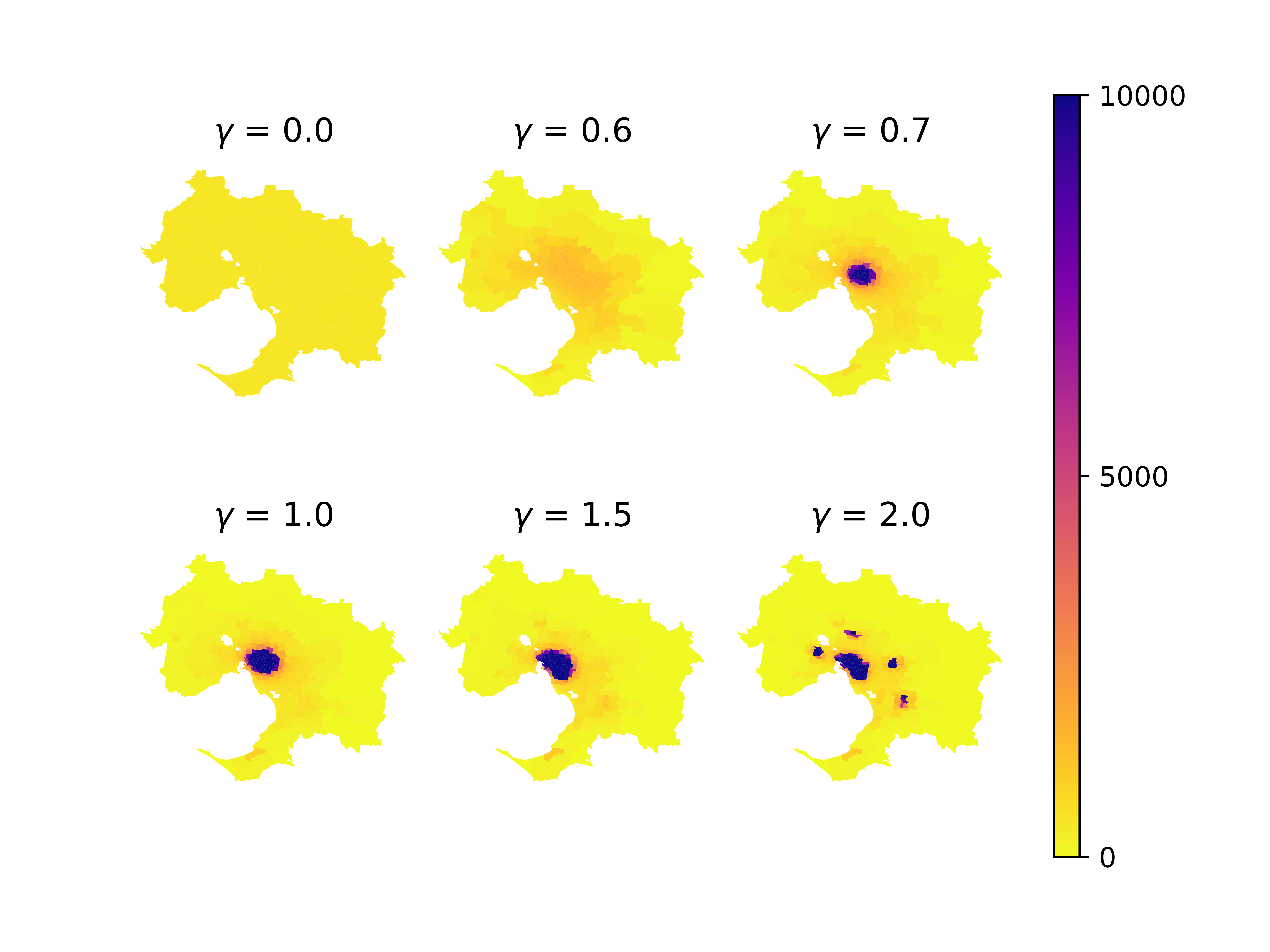}
\caption{Abrupt transition between uniform, monocentric and polycentric phases in Melbourne: population density maps for different values of $\gamma$. Value of $\alpha$ is fixed at the current level. Values of $\gamma$ are shown in relative units: divided by the estimated current value.}
\label{fig:Melbourne_sharp}
\end{figure}

\begin{figure}
\centering
\includegraphics[width=.9\linewidth]{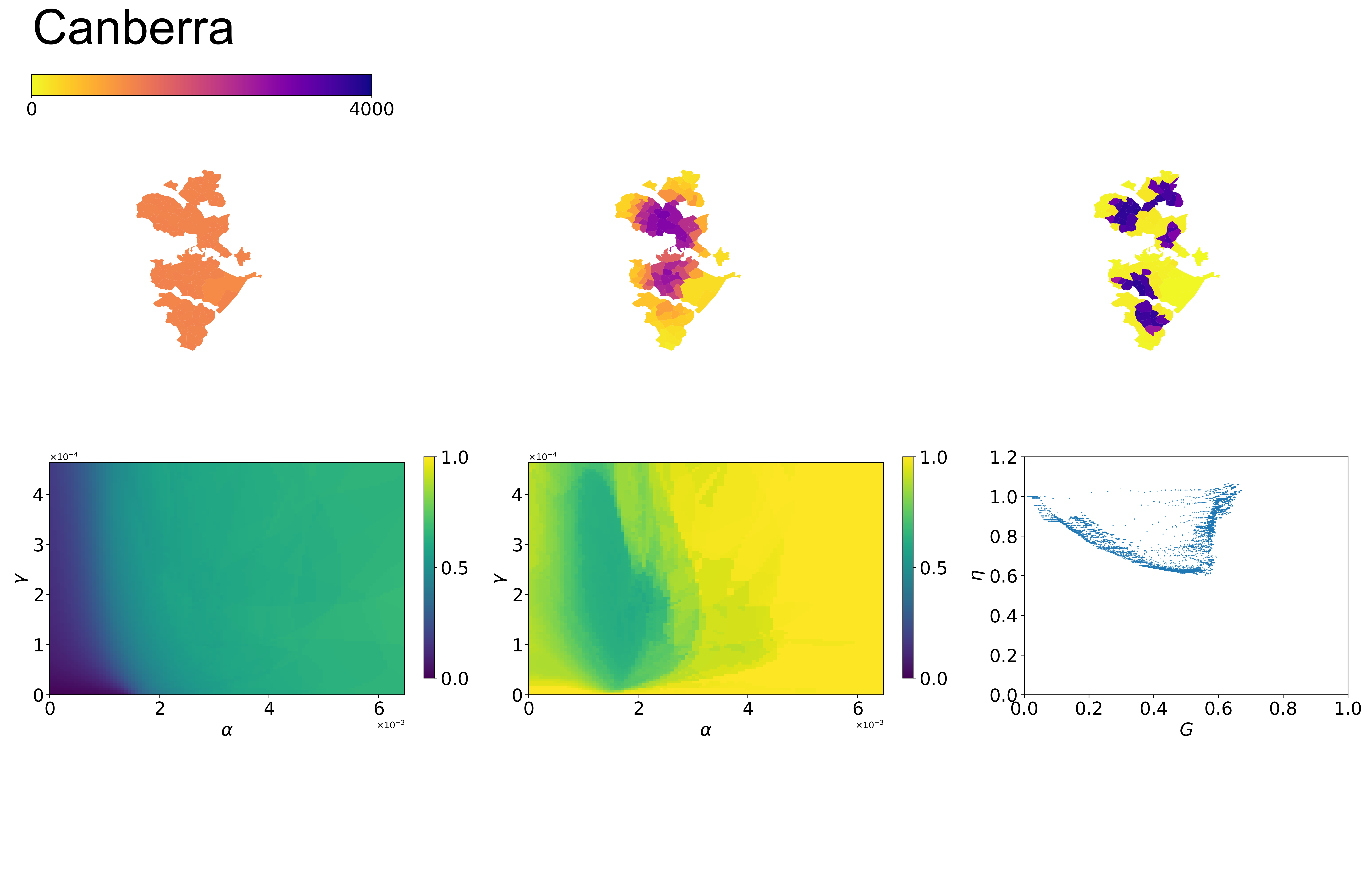}
\caption{Equilibrium phases of Canberra. In the first row, we show the population density maps for three distinct phases of equilibrium city structure: uniform (left), monocentric (middle) and polycentric (right).   In the second row, the values of $G$ and $\eta$ near equilibrium are displayed. The left figure displays value of $G$ is shown as a function of $\alpha$ and $\gamma$, the middle one displays the corresponding values for $\eta$, the right figure displays values of $G$ and $\eta$ plotted against each other. Values of the population density are measured in thousands of residents per $km^2$.}
\label{fig:Canberra_exception}
\end{figure}

\begin{figure}
\centering
\includegraphics[width=\textwidth]{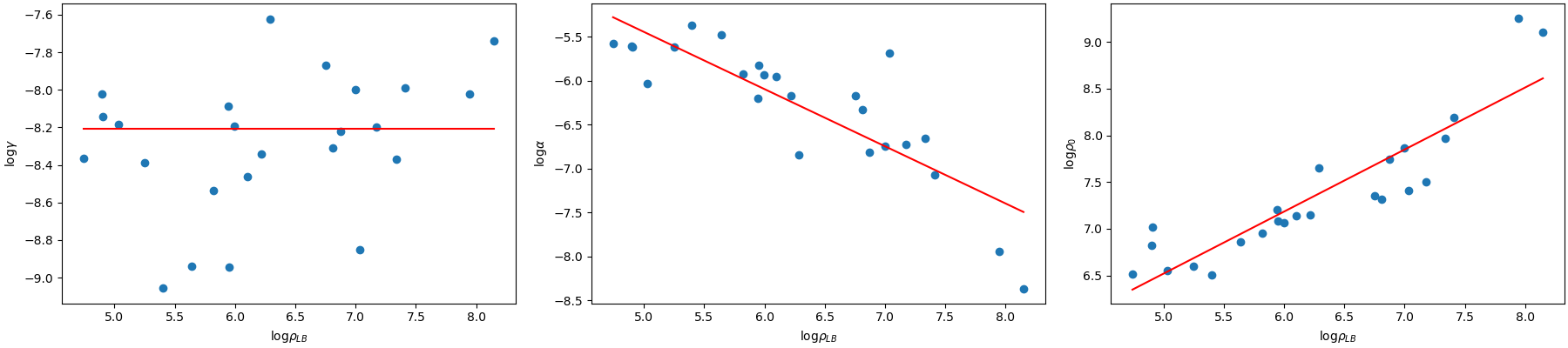}
\caption{Logarithmic values of parameters $\gamma$, $\alpha$ and $\rho_0$ plotted against LouBar density ($\rho_{LB}$) logarithm.}
\end{figure}

\begin{figure}
    \centering
    \includegraphics[width=\textwidth]{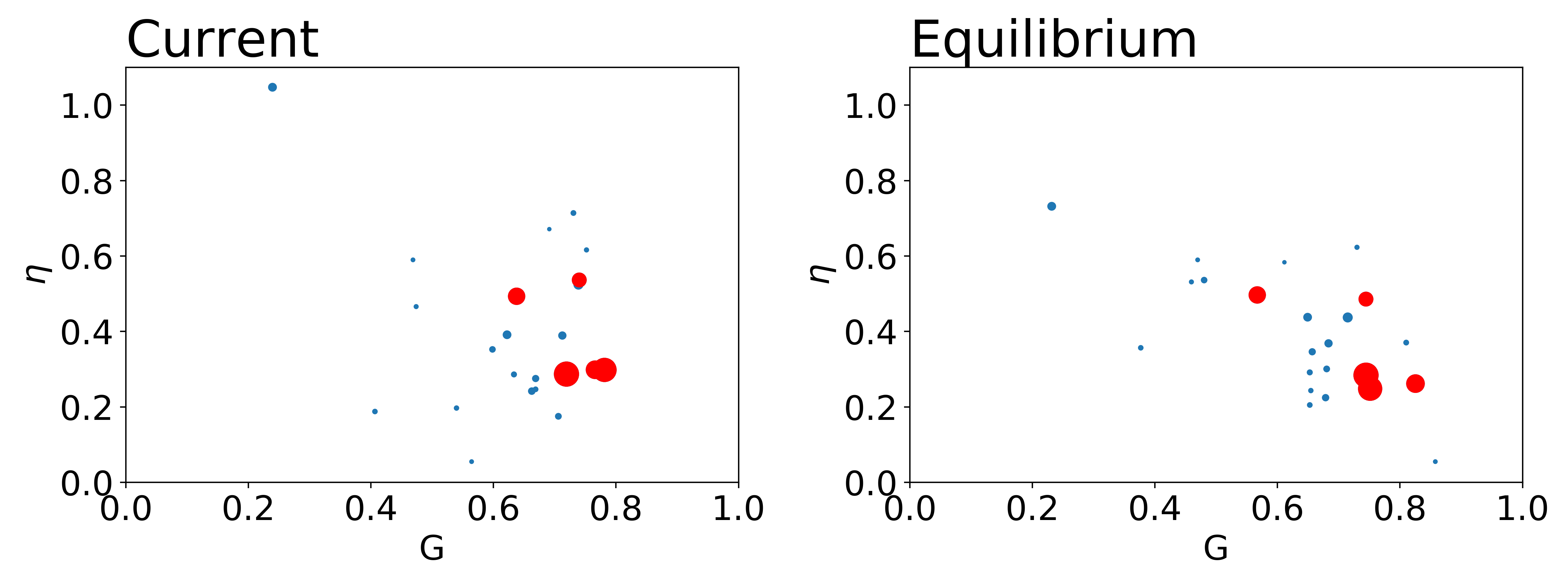}
    \caption{Arrangement of the Australian urban areas in the heterogeneity--spreading space based on 2011 Census data: current and near equilibrium. The size of each circle corresponds to the size of the corresponding urban area; urban areas with population above 1 million people are shown in red. }
    \label{fig:spreading_heterogeneity_diagram_2011}
\end{figure}

\begin{table}[h!]
\centering
    \caption{Accuracy of the calibrated resettlement model: values of the Spearman's rank correlation coefficient, the corresponding p-value. The values are calculated for both 2011 and 2015 five-year migration datasets.}
    \vspace{3mm}
    \label{table:calibration_precision}
    \begin{tabular}{lrrrrr}
        \multirow{2}{*}{GCA} &
        \multirow{2}{*}{Sample size} &
        \multicolumn{2}{c}{2016}  & \multicolumn{2}{c}{2011} \\
        && Spearman $r$ 
        & p-value
        & Spearman $r$ 
        & p-value \\
        \midrule
		Sydney & 87025 & 0.49 & $<10^{-308}$ & 0.51 & $<10^{-308}$ \\
        Melbourne & 91809 & 0.49 & $<10^{-308}$ & 0.48 & $<10^{-308}$ \\
        Brisbane & 50176 & 0.49 & $<10^{-308}$ & 0.47 & $<10^{-308}$ \\
        Adelaide & 10816 & 0.59 & $<10^{-308}$ & 0.57 & $<10^{-308}$ \\
        Perth & 22201 & 0.56 & $<10^{-308}$ & 0.55 & $<10^{-308}$ \\
        Hobart & 1156 & 0.35 & 5.9$\times10^{-35}$ & 0.32 & 3.2$\times10^{-29}$ \\
        Darwin & 1369 & 0.26 & 5.1$\times10^{-23}$ & 0.29 & 2.6$\times10^{-30}$ \\
        Canberra & 10816 & 0.56 & $<10^{-308}$ & 0.59 & $<10^{-308}$ \\
        GoldCoast & 2304 & 0.56 & 3.9$\times10^{-192}$ & 0.54 & 1.1$\times10^{-157}$ \\
        Newcastle & 841 & 0.56 & 9.0$\times10^{-71}$ & 0.57 & 3.1$\times10^{-73}$ \\
        SunshineCoast & 1089 & 0.66 & 4.6$\times10^{-139}$ & 0.60 & 4.9$\times10^{-96}$ \\
        Illawarra & 484 & 0.7 & 1.4$\times10^{-73}$ & 0.73 & 8.4$\times10^{-49}$ \\
        Geelong & 324 & 0.67 & 5.0$\times10^{-43}$ & 0.62 & 7.6$\times10^{-32}$ \\
        Townsville & 529 & 0.38 & 2.0$\times10^{-19}$ & 0.28 & 1.2$\times10^{-10}$ \\
        Cairns & 576 & 0.53 & 1.1$\times10^{-42}$ & 0.53 & 1.5$\times10^{-43}$ \\
        Toowoomba & 196 & 0.55 & 5.4$\times10^{-17}$ & 0.49 & 2.2$\times10^{-13}$ \\
        Ballarat & 81 & 0.79 & 3.1$\times10^{-18}$ & 0.81 & 3.1$\times10^{-20}$ \\
        Bendigo & 121 & 0.84 & 6.8$\times10^{-33}$ & 0.85 & 1.3$\times10^{-34}$ \\
        Mackay & 169 & 0.57 & 3.7$\times10^{-16}$ & 0.60 & 6.1$\times10^{-16}$ \\
        Launceston & 289 & 0.32 & 1.9$\times10^{-8}$ & 0.26 & 7.2$\times10^{-06}$ \\
        Bunbury & 144 & 0.76 & 1.1$\times10^{-28}$ & 0.88 & 1.5$\times10^{-12}$ \\
        Rockhampton & 169 & 0.52 & 2.9$\times10^{-13}$ & 0.52 & 7.3$\times10^{-13}$ \\
        HerveyBay & 36 & 0.56 & 3.0$\times10^{-4}$ & 0.48 & 3.4$\times10^{-3}$ \\
        Bundaberg & 81 & 0.41 & 1.4$\times10^{-4}$ & 0.39 & 3.2$\times10^{-4}$ \\
        \bottomrule
    \end{tabular}
\end{table}

\end{document}